\documentclass[preprint,amsmath,amssymb]{revtex4}
\usepackage{amsmath}
\usepackage{xcolor}
\usepackage{pstricks}
\usepackage{color}
\usepackage{graphicx}
\usepackage{latexsym} % Gets \Box etc
\usepackage{amssymb}  % \gtrsim, \geqslant, etc etc: see amsguide.ps
\usepackage{amsbsy}
\newcommand{\al}{\alpha}
\newcommand{\be}{\beta}
\newcommand{\ga}{\gamma}
\newcommand{\Ga}{\Gamma}
\newcommand{\de}{\delta}

\newcommand{\ka}{\kappa}
\newcommand{\la}{\lambda}

\newcommand{\del}{\nabla}
\newcommand{\si}{\sigma}

\renewcommand{\th}{\theta}   % LaTeX: \th already defined

\newcommand{\Om}{\Omega}

\newcommand{\p}{\partial}
\newcommand{\<}{\langle}
\renewcommand{\>}{\rangle} % LaTeX: \> already defined

\newcommand{\beq}{\begin{equation}}
\newcommand{\eeq}{\end{equation}}
\newcommand{\ba}{\begin{array}}
\newcommand{\bea}{\begin{eqnarray}}
\newcommand{\ea}{\end{array}}
\newcommand{\eea}{\end{eqnarray}}
\newcommand{\bi}{\begin{itemize}}  %\setlength{\itemsep}{0\parsep}}
\newcommand{\ei}{\end{itemize}}
\newcommand{\ben}{\begin{enumerate}} %\setlength{\itemsep}{0\parsep}}
\newcommand{\een}{\end{enumerate}}
\newcommand{\bc}{\begin{center}}
\newcommand{\ec}{\end{center}}
\newcommand{\bl}{\begin{flushleft}}
\newcommand{\el}{\end{flushleft}}
\newcommand{\br}{\begin{flushright}}
\newcommand{\er}{\end{flushright}}
\newcommand{\fzero}{f^{(0)}}
\newcommand{\fone}{f^{(1)}}
\newcommand{\ftwo}{f^{(2)}}

\newcommand\comment[1]{ \hbox{[{\it Comment suppressed here.}\/]} }
\newcommand\hide[1]{}

\newcommand{\dif}{\mathrm{d}}

\newcommand{\skipover}[1]{}
\newcommand{\nn}{\nonumber \\}

\renewcommand{\l}{\left}
\renewcommand{\r}{\right}

\def\lsim{\mathrel{\rlap{\lower4pt\hbox{\hskip1pt$\sim$}}
    \raise1pt\hbox{$<$}}}
\def\gsim{\mathrel{\rlap{\lower4pt\hbox{\hskip1pt$\sim$}}
    \raise1pt\hbox{$>$}}}
\usepackage{graphicx}

\newcommand{\lie}{\mathcal{L}_{\xi}}

\begin{document}

\title{Conformal symmetry and non-relativistic second order fluid dynamics}

\author{Jingyi Chao and Thomas Sch\"afer}

\address{Physics Department, North Carolina State University, 
 Raleigh, NC 27695, USA}

\date{\today}

\begin{titlepage}
\renewcommand{\thepage}{}          % No page number on title page

\begin{abstract}
We study the constraints imposed by conformal symmetry on the 
equations of fluid dynamics at second order in gradients of the 
hydrodynamic variables. At zeroth order conformal symmetry 
implies a constraint on the equation of state, ${\cal E}_0=\frac{2}{3}
P$, where ${\cal E}_0$ is the energy density and $P$ is the pressure. 
At first order, conformal symmetry implies that the bulk viscosity 
must vanish. We show that at second order conformal invariance 
requires that two-derivative terms in the stress tensor must be
traceless, and that it determines the relaxation of dissipative 
stresses to the Navier-Stokes form. We verify these results 
by solving the Boltzmann equation at second order in the gradient 
expansion. We find that only a subset of the terms allowed by 
conformal symmetry appear.
\end{abstract}

\maketitle
\end{titlepage}

%%%%%%%%%%%%%%%%%%%%%%%%%%%%%%%%%%%%%%%%%%%%%%%%%%%%%%%%%%%%%%%%%%%%%%%%%%%
\section{Introduction}
\label{sec_intro}
%%%%%%%%%%%%%%%%%%%%%%%%%%%%%%%%%%%%%%%%%%%%%%%%%%%%%%%%%%%%%%%%%%%%%%%%%%%

 There has been great interest recently in the dynamics of conformally 
invariant fluids. One motivation is the observation that the AdS/CFT 
correspondence \cite{Maldacena:1997re}, which relates the dynamics 
of a four-dimensional conformal field theory to quantum gravity in 
ten dimensions, also implies a fluid-gravity correspondence 
\cite{Kovtun:2003wp,Bhattacharyya:2008jc,Rangamani:2009xk}. The 
fluid-gravity correspondence relates solutions of the Navier-Stokes
equation for conformally invariant fluids in four dimensions to dynamical 
black hole solutions of the Einstein equations in higher dimensions. The 
fluid-gravity correspondence has been used to establish the form of second 
order terms in the equations of relativistic fluid dynamics 
\cite{Baier:2007ix,Bhattacharyya:2008jc}. The correspondence has also 
been extended to certain non-conformal fluids \cite{Kanitscheider:2009as}, 
superfluids \cite{Herzog:2011ec,Bhattacharya:2011tr} , and to the 
incompressible non-relativistic Navier-Stokes equation \cite{Fouxon:2008tb}.

 Another motivation comes from the experimental discovery of nearly 
perfect fluidity in certain conformal or almost conformal fluids 
\cite{Schafer:2009dj}. The first example is the observation of nearly 
ideal flow of the quark gluon plasma produced in heavy ion collisions 
at the relativistic heavy ion collider (RHIC) \cite{Adler:2003kt}. This 
discovery was soon followed by the discovery of nearly ideal fluid 
dynamics in a very different system, the dilute Fermi gas at unitarity 
\cite{oHara:2002}. Both the quark gluon plasma and the dilute Fermi 
gas are characterized by a very small shear viscosity to entropy density
ratio, $\eta/s\lsim 0.4\hbar/k_B$ \cite{Dusling:2007gi,Romatschke:2007mq,Song:2007ux,Schafer:2007pr,Turlapov:2007,Schaefer:2009px,Cao:2010wa}. 
This result is close to the value $\eta/s=\hbar/(4\pi k_B)$ obtained
in the strong coupling limit of a super-conformal Yang Mills plasma
\cite{Policastro:2001yc,Kovtun:2004de}.

 In the present work we will focus on the dilute Fermi gas and study 
the hydrodynamics of a scale invariant non-relativistic fluid. At unitarity 
the two-body scattering length of the fermions is infinite and the range of 
the interaction is zero. This implies that the system is exactly scale 
invariant \cite{Son:2005rv,Mehen:1999nd,Hagen:1972pd}. Hydrodynamics is 
an effective theory that governs the behavior of the system at long 
distances. The equations of motion are organized as an expansion in the 
number of derivatives acting on the hydrodynamic variables. In the case 
of a one-component non-relativistic fluid the hydrodynamic variables are 
the temperature $T$, the density $n$, and the fluid velocity $\vec{V}$. 
At zeroth order in the gradient expansion conformal symmetry implies a 
constraint on the equation of state, ${\cal E}_0=\frac{2}{3}P$, where 
${\cal E}_0(n,T)$ is the energy density and $P(n,T)$ is the pressure. 
At first order, conformal symmetry implies that the contribution of 
the divergence of the velocity field to the pressure, $\delta P=\zeta 
(\nabla\cdot V)$, must vanish \cite{Son:2005tj,Enss:2010qh,Castin:2011}. 
This means that the bulk viscosity $\zeta$ of a scale invariant fluid is 
zero. In this work we will study constraints from conformality at second 
order in the gradient expansion. 

 Second order hydrodynamics has a long history \cite{Burnett:1935,Israel:1979wp,Garcia:2008} but the theory is rarely used in practice. 
An important exception is the relativistic case, where the second 
order theory has the advantage of being manifestly causal 
\cite{Romatschke:2009im}. There is a similar reason for using 
second order hydrodynamics in certain non-relativistic flows.
The expansion of a scale invariant fluid after the release from 
a harmonic trap leads to a Hubble-like flow in which the expansion
velocity is linear in the coordinates \cite{Schaefer:2009px}. This 
implies that ideal stresses propagate outward with a finite 
velocity, but at first order in gradients dissipative stresses 
build up instantaneously everywhere in space. This unphysical 
behavior can be avoided by taking into account a finite relaxation
time for the dissipative stresses. We will show that relaxation
time effects appear naturally at second order in the gradient 
expansion. 

 There are additional reasons for studying higher order terms in the 
equations of fluid dynamics. Higher order terms provide information 
about the convergence of the gradient expansion, and they lead to new 
Kubo relations. In the relativistic case it is known that some of 
these Kubo formulas relate transport properties to thermodynamic 
quantities \cite{Moore:2010bu}. Finally, understanding the structure
of the gradient expansion may help in constructing holographic 
duals of non-relativistic conformal field theories 
\cite{Son:2008ye,Balasubramanian:2008dm,Herzog:2008wg}. The paper 
is structured as follows. In Sect.~\ref{sec_conf_hyd} we review a 
local extension of Galilean and conformal invariance and study 
constraints on the structure of gradient terms in fluid dynamics. 
In Sect.~\ref{sec_kin} we consider a specific kinetic equation 
and solve for the stress tensor and the entropy current at 
second order in gradients. We show that the result agrees with 
the general constraints, but that the kinetic theory only generates
a subset of the allowed terms. We conclude with a discussion 
of open questions. 

%%%%%%%%%%%%%%%%%%%%%%%%%%%%%%%%%%%%%%%%%%%%%%%%%%%%%%%%%%%%%%%%%%%%%%%%%%%
\section{Conformal hydrodynamics}
\label{sec_conf_hyd}
%%%%%%%%%%%%%%%%%%%%%%%%%%%%%%%%%%%%%%%%%%%%%%%%%%%%%%%%%%%%%%%%%%%%%%%%%%%

%%%%%%%%%%%%%%%%%%%%%%%%%%%%%%%%%%%%%%%%%%%%%%%%%%%%%%%%%%%%%%%%%%%%%%%%%%%
\subsection{Navier-Stokes equation}
\label{sec_ns}
%%%%%%%%%%%%%%%%%%%%%%%%%%%%%%%%%%%%%%%%%%%%%%%%%%%%%%%%%%%%%%%%%%%%%%%%%%%

 The constraints imposed by Galilean invariance and conformal symmetry 
on the structure of the Navier-Stokes equation were studied by Son 
\cite{Son:2005tj} based on earlier work by Son and Wingate \cite{Son:2005rv}. 
The basic strategy is to generalize Galilean and scale invariance 
to local symmetries. For this purpose we consider the fluid moving 
in a curved background characterized by the metric $g_{ij}(x,t)$. 
We also include background $U(1)$ gauge fields $A_0(x,t)$ and 
$A_i(x,t)$. The Navier-Stokes equations in curved space are
\bea
\label{h_cont}
\frac{1}{\sqrt{g}}\partial_{t}\left(\sqrt{g}\rho\right)
   +\nabla_{i}\left(\rho V^{i}\right)  &=& 0\, ,  \\
\label{h_mom}
\frac{1}{\sqrt{g}}\partial_{t}\left(\sqrt{g}\rho V_{i}\right)
   +\nabla_{k}\Pi^{k}_{\, i}  &=& \frac{\rho}{m}\,Q_i \, ,  \\
\label{h_entr}
\frac{1}{\sqrt{g}}\partial_{t}\left(\sqrt{g}s\right)
   +\nabla_{k} \jmath_s^k
   &=& \frac{R}{T} \, ,
\eea
where $g=\det(g_{ij})$ and $\nabla_i$ is the covariant derivative 
associated with the metric $g_{ij}$. The hydrodynamic variables
are the mass density $\rho$, the entropy density $s$, and the fluid 
velocity $V^i$. The energy momentum tensor is $\Pi_{ij} = \Pi_{ij}^0
+\delta\Pi_{ij}$, where
\beq 
\Pi_{ij}^0 = \rho V_i V_j +P g_{ij} 
\eeq
is the ideal fluid part, and $\delta\Pi_{ij}$ is the viscous
correction. The entropy current is $\jmath_s^i=sV^i+\delta
\jmath_s^i$, and $R$ is the dissipative function. We will specify 
the viscous terms $\delta\Pi_{ij},\delta\jmath_s^i$ and $R$ below. 
The Lorentz force $Q_i$ is defined as $Q_i=E_i-F_{ij}V^j$ with 
$E_i=\p_t A_i-\p_iA_0$ and $F_{ij}=\p_i A_j-\p_jA_i$.

%%%%%%%%%%%%%%%%%%%%%%%%%%%%%%%%%%%%%%%%%%%%%%%%%%%%%%%%%%%%%%%%%%%%%%%%%%%
\subsection{Diffeomorphism invariance}
\label{sec_diff}
%%%%%%%%%%%%%%%%%%%%%%%%%%%%%%%%%%%%%%%%%%%%%%%%%%%%%%%%%%%%%%%%%%%%%%%%%%%

 The equations of fluid dynamics are invariant under a combination of 
local diffeomorphisms $x^i\to x^i+\xi^i(x,t)$ and gauge transformations
$A_0\to A_0-\dot\alpha$, $A_i\to A_i-\partial_i\alpha$. This symmetry
generalizes Galilean invariance. A Galilei transformation with boost 
velocity $U^i$ corresponds to a flat background metric $g_{ij}=\delta_{ij}$ 
and $\xi^i= U^it$ as well as $\alpha=mU^ix^i$. The hydrodynamic
variables are invariant under gauge transformations. They transform
under diffeomorphisms as 
\bea 
 \de \rho  &=& -\xi^{k}\nabla_k \rho \, , \\
 \de s     &=& -\xi^{k}\nabla_k s\, , \\
 \de V^{i} &=& -\xi^{k}\nabla_{k}V^{i}+V^{k}\nabla_{k}\xi^{i}
               +\dot{\xi}^{i} \, .  
\eea
We note that $\rho$ and $s$ transform as scalars, but $V_i$ does
not transform as a vector. We have $\de V^i= {\cal L}_\xi V^i 
+\dot\xi^i$, where ${\cal L}_\xi$ is the Lie derivative and 
$\dot\xi^i$ is an ``extra'' term. The gauge fields and the metric 
transform as 
\bea
  \de A_{0} &=& -\xi^{k}\nabla_{k}A_{0} - A_{k}\dot{\xi}^{k} \, , \\ 
  \de A_{i} &=& -\xi^{k}\nabla_{k}A_{i} - A_{k}\nabla_{i}\xi^{k}
                 +mg_{ik}\dot{\xi}^{k}\, ,  \\
  \de g_{ij}&=& -g_{ik}\del_{j}\xi^{k} -g_{kj}\del_{i}\xi^{k}\, .
\end{eqnarray}
Using the transformation laws for the gauge fields we can deduce the 
transformation properties of the field strengths and the Lorentz force.
We find  
\bea
  \de E_{i} &=& \lie E_{i} + F_{ik}\dot{\xi}^{k}
    + m\dot{g}_{ik}\dot{\xi}^{k} + mg_{ik}\ddot{\xi}^{k}\, , \\
  \de F_{ij} &=& \lie F_{ij} + mg_{jk}\del_{i}\dot{\xi}^{k}
    - mg_{ik}\del_{j}\dot{\xi}^{k} \, , \\
  \de Q_i   &=&   \lie Q_i  + m\left( 
      \dot{g}_{ik}\dot{\xi}^{k} + g_{ik}\ddot{\xi}^{k}
    + V_{k}\del_{i}\dot{\xi}^{k}-g_{ik}V^{j}\del_{j}\dot{\xi}^{k}\right) \, . 
\eea
Using these ingredients we can verify that the hydrodynamic equations 
(\ref{h_cont}-\ref{h_entr}) are invariant. We note that under diffeomorphisms 
the ideal parts of the momentum and entropy currents do not transform as 
vectors, and the ideal part of the stress tensor does not transform as a 
tensor. The extra terms in the transformation laws are canceled by terms
that arise from the time derivatives of the conserved charges and the 
Lorentz force. In order to maintain diffeomorphism invariance beyond
the level of ideal hydrodynamics dissipative corrections to the 
entropy current and the stress tensor have to transform as vectors 
and second rank tensors, respectively. 

 At first order in gradients dissipative terms involve the heat current
and the shear and bulk stresses \cite{Landau:fl}. The heat current 
\beq 
 q_i =-\nabla_i \log(T)\, , 
\eeq
where $\nabla_i$ is the covariant derivative, transforms as a vector under 
diffeomorphisms. The shear and bulk stresses can be promoted to second rank 
tensors by adding extra terms involving $\dot g_{ij}$ \cite{Son:2005tj}. We 
will show below that these terms automatically appear in kinetic theory. We 
have
\bea
\label{def_sig}
 \si_{ij} &=& \del_{i}V_{j}+\del_{j}V_{i}+\dot{g}_{ij}
         -\frac{2}{3}g_{ij}\langle\sigma\rangle \, , \\
\label{def_s}
 \langle\si\rangle &=& \del\cdot V+\frac{\dot{g}}{2g}\, 
\eea
where $\si_{ij}$ is the shear stress tensor and $g_{ij}\langle\sigma
\rangle$ is the bulk stress tensor. 

%%%%%%%%%%%%%%%%%%%%%%%%%%%%%%%%%%%%%%%%%%%%%%%%%%%%%%%%%%%%%%%%%%%%%%%%%%%
\subsection{Conformal invariance}
\label{sec_conf}
%%%%%%%%%%%%%%%%%%%%%%%%%%%%%%%%%%%%%%%%%%%%%%%%%%%%%%%%%%%%%%%%%%%%%%%%%%%

 Conformal transformations are infinitesimal rescalings of the time 
variable, $t\to t+\beta(t)$. A non-relativistic scale transformation 
corresponds to $\beta=bt$ combined with a diffeomorphism generated
by $\xi^k=\frac{1}{2}bx^k$. Special conformal transformations are 
generated by $\beta=-ct^2$, $\xi^k=-ctx^k$ and $\alpha =-\frac{1}{2}
mcx^2$. An operator $O$ is said to have conformal dimension $\Delta_O=
[O]$ if
\beq
 \delta O = -\beta \dot O - \Delta_O\dot\beta O\, .
\eeq
Examples of operators with well-defined conformal dimension are 
\beq 
[g_{ij}]=-1\, , \hspace{0.5cm}
[g^{ij}]=+1\, , \hspace{0.5cm}
[A_0]=+1   \, , \hspace{0.5cm}
[A_i]=0\, . 
\eeq
We also have $[O_1O_2]=[O_1]+[O_2]$ and $[\partial_i O]=[O]$, 
but $\partial_t O$ does not, in general, have a well-defined 
conformal dimension. The conformal dimensions of the hydrodynamic
variables are
\beq
[\rho]=\frac{3}{2}\, , \hspace{0.5cm}
[s]=\frac{3}{2}\, , \hspace{0.5cm}
[V_i]=0   \, .
\eeq
We can now check under what conditions the equations of hydrodynamics 
are conformally invariant. Two key relations are
\beq
\label{del_dotgij}
 \delta\dot g_{ij} = -\beta \ddot g_{ij} + \ddot \beta g_{ij}
     \equiv \delta^{\Delta=0}( \dot g_{ij})  + \ddot \beta g_{ij}\, , 
\eeq
and 
\beq
\label{del_dotg_g}
 \delta \left(\frac{\dot{g}}{2g}\right) = 
  -\beta \partial_t \left(\frac{\dot{g}}{2g}\right)
  -\dot\beta \left(\frac{\dot{g}}{2g}\right) + \frac{3}{2}\ddot \beta
  =  \delta^{\Delta=1}\left(\frac{\dot{g}}{2g}\right) 
  + \frac{3}{2}\ddot \beta\, , 
\eeq
where $\delta^{\Delta=d}O$ denotes the transformation of a conformal
operator $O$ of dimension $d$. Using equ.~(\ref{del_dotgij}) and 
(\ref{del_dotg_g}) we can show that the equations of fluid dynamics
are conformally invariant if 
\beq 
[\Pi_{ij}]= \frac{3}{2}\, , \hspace{0.5cm}
[(\jmath_s)_i]= \frac{3}{2}\, , \hspace{0.5cm}
[R]= \frac{7}{2}\, .
\eeq
These relations are satisfied in the case of ideal fluid dynamics
(in ideal hydrodynamics $R=0$). At one derivative order we find 
$[\sigma_{ij}]= 0$, $[q_i]=0$ and $ \delta \langle\sigma\rangle   
= \delta^{\Delta=1}\langle\sigma\rangle + \frac{3}{2}\ddot \beta$. 
This implies that 
\beq
\label{del_j_conf}
 \delta\Pi_{ij}=-\eta\sigma_{ij}\, , \hspace{0.5cm}
 \delta(\jmath_s)_i = \kappa q_i \, , 
\eeq
are conformally invariant provided  $[\eta]=[\kappa]=\frac{3}{2}$,
but $\delta\Pi_{ij}=-\zeta g_{ij}\langle \sigma\rangle$ violates
conformal invariance unless $\zeta=0$. We note that the conformal
dimension of $\eta$ and $\kappa$ agree with their naive scaling 
dimension.

%%%%%%%%%%%%%%%%%%%%%%%%%%%%%%%%%%%%%%%%%%%%%%%%%%%%%%%%%%%%%%%%%%%%%%%%%%%
\subsection{Second order terms: Building blocks}
\label{sec_inv}
%%%%%%%%%%%%%%%%%%%%%%%%%%%%%%%%%%%%%%%%%%%%%%%%%%%%%%%%%%%%%%%%%%%%%%%%%%%

 In order to construct the most general second order terms we begin 
by listing conformally invariant building blocks, that is diffeomorphism
invariant scalars, vectors, and tensors with well-defined conformal
dimensions. These objects are constructed from the hydrodynamic 
variables $T,P$ and $V^i$, the metric $g_{ij}$ and the gauge fields 
$A_0,A_i$, and time or spatial derivatives. Instead of $T$ and $P$ 
it is sometimes useful to consider $T$ and the dimensionless variable 
$\alpha=\mu/T$, where $\mu$ is the chemical potential, as independent
quantities. We begin with one-derivative scalars. Diffeomorphism invariant 
one-derivative scalars constructed from the hydrodynamic variables are $DT$, 
$DP$ and $\langle\sigma\rangle$, where $D=\partial_0+V^k\nabla_k$ is the 
comoving derivative. Because there are two scalar equations of motion
only one of these objects is linearly independent. Neither one of 
the one-derivative scalars has a well-defined conformal dimension
but we can construct conformal scalars by taking linear combinations. 
The quantity
\beq 
{\cal S} = \left(D+\frac{2}{3}\langle\sigma\rangle\right)T
\eeq
is a conformal scalar with dimension $[{\cal S}]=2$. We will show,
however, that at leading order in the derivative expansion ${\cal S}$ 
vanishes by the equations of motion. We can construct three conformal 
vectors
\beq
{\cal V}^1_i=\nabla_i T\, ,\hspace{0.5cm}
{\cal V}^2_i=\nabla_i P\, ,\hspace{0.5cm}
{\cal V}^3_i=DV_i-\frac{Q_i}{m}\, ,
\eeq
with weights $[{\cal V}^1_i]=1$, $[{\cal V}^2_i]=\frac{5}{2}$, and 
$[{\cal V}^3_i]=1$. There is one vector equation of motion and, up
to higher order terms in the derivative expansion, ${\cal V}^3$ can 
be expressed in terms of ${\cal V}^2$. There are two one-derivative
tensors. The first is the shear tensor defined in equ.~(\ref{def_sig}).
The second is related to the vorticity of the fluid. We have
\beq
{\cal T}^1_{ij} = \sigma_{ij}\, ,\hspace{0.5cm}
{\cal T}^2_{ij} \equiv \Omega_{ij}
  = \nabla_iV_j-\nabla_jV_i-\frac{F_{ij}}{m}\, ,
\eeq
with $[{\cal T}^1_{ij}]=[{\cal T}^2_{ij}]=0$.

%%%%%%%%%%%%%%%%%%%%%%%%%%%%%%%%%%%%%%%%%%%%%%%%%%%%%%%%%%%%%%%%%%%%%%%%%%%
\subsection{Second order terms: Stress tensor}
\label{sec_delpi_2}
%%%%%%%%%%%%%%%%%%%%%%%%%%%%%%%%%%%%%%%%%%%%%%%%%%%%%%%%%%%%%%%%%%%%%%%%%%%

 In this section we list conformal two-derivative tensors that 
can contribute to the stress tensor in a conformal theory. Rotational 
invariance implies that the stress tensor has to be symmetric. We will 
argue below that in a conformal theory the dissipative part of the stress
tensor also has to be traceless. One class of two-derivative tensors
arises from contracting two one-derivative tensors. We find
\beq
\label{O_ij_123}
{\cal O}_{ij}^1 = \sigma_{\langle i}^{\;\;\; k}\sigma^{}_{j\rangle k}\, , 
  \hspace{0.5cm}
{\cal O}_{ij}^2 = \sigma_{\langle i}^{\;\;\; k}\Omega^{}_{j\rangle k}\, ,
  \hspace{0.5cm}
{\cal O}_{ij}^3 = \Omega_{\langle i}^{\;\;\; k}\Omega^{}_{j\rangle k}\, . 
\eeq
with $[{\cal O}^1_{ij}]=[{\cal O}^2_{ij}]=[{\cal O}^3_{ij}]=1$. Here, 
${\cal O}_{\langle ij\rangle}=\frac{1}{2}({\cal O}_{ij}+{\cal O}_{ji}
-\frac{2}{3}g_{ij}{\cal O}^k_{\;\;k})$ denotes the symmetric traceless 
part of ${\cal O}_{ij}$. There is a unique diffeomorphism invariant
tensor of conformal dimension $+1$ that can be constructed from the 
comoving derivative of $\sigma_{ij}$. This tensor is given by
\beq
\label{O_ij_4}
{\cal O}_{ij}^4 = g_{ik}\dot\sigma^{k}_{\; j} + V^k\nabla_k \sigma_{ij}
  + \frac{2}{3} \langle \sigma\rangle \sigma_{ij} 
  + F_{( i}^{\;\;\; k}\sigma^{}_{kj)} \, ,
\eeq
where ${\cal O}_{(ij)}=\frac{1}{2}({\cal O}_{ij}+{\cal O}_{ji})$. 
There is a set of conformal tensors that can be obtained as the 
tensor product of two one-derivative tensors, or as two covariant
derivatives acting on a scalar. We find 
\bea
\label{O_ij_56789}
&{\cal O}_{ij}^5 = \nabla_{\langle i}T\nabla_{j\rangle}T\, , 
  \hspace{0.5cm}
{\cal O}_{ij}^6 = \nabla_{\langle i}P\nabla_{j\rangle}P\, ,
  \hspace{0.5cm}
{\cal O}_{ij}^7 = \nabla_{\langle i}T\nabla_{j\rangle}P\, , &\\ 
&{\cal O}_{ij}^8 = \nabla_{\langle i}\nabla_{j\rangle}T\, , 
  \hspace{0.5cm}
{\cal O}_{ij}^9 = \nabla_{\langle i}\nabla_{j\rangle}P\, ,&
\eea
with $[{\cal O}^{5-9}_{ij}]=\{2,5,\frac{7}{2},1,\frac{5}{2}\}$. Finally, 
in curved space there is a two-derivative tensor constructed only 
from the metric. This is the traceless part of the Ricci tensor, 
\beq 
\label{O_ij_10}
{\cal O}_{ij}^{10} = R_{\langle ij\rangle}\, . 
\eeq
We can compare this list with the analogous result in the relativistic
case \cite{Baier:2007ix}. In relativistic fluid dynamics there are five
second order terms (four in flat space). Three of those are analogous
to ${\cal O}^1_{ij}-{\cal O}^3_{ij}$, and one is similar to ${\cal O}^4_{ij}$.
It is interesting to note that the relativistic analog of 
equ.~(\ref{O_ij_4}) has a coefficient $\frac{1}{3}$ in front of $\langle
\sigma\rangle$ instead of $\frac{2}{3}$. This difference is similar 
to the difference between the relativistic and non-relativistic 
conformal equation of state, ${\cal E}_0=\frac{1}{3}P$ versus
${\cal E}_0=\frac{2}{3}P$. In a relativistic theory without conserved 
charges there are only two hydrodynamic variables, $T$ and the 
four-velocity $U^\mu$,  and as a result there are no terms analogous 
to ${\cal O}^6_{ij}, {\cal O}^7_{ij}$ and ${\cal O}^9_{ij}$. Finally, 
in the relativistic case there is only one linear combination of 
${\cal O}^5_{ij}$ and  ${\cal O}^8_{ij}$ which is conformal, and that 
linear combination can be related to ${\cal O}^4_{ij}$ using the 
equations of motion. We can also compare our result to the structure 
of the Burnett equations \cite{Garcia:2008}. The Burnett equations 
contain six independent two-derivative terms in $\Pi_{ij}$. The reason 
that the number of independent terms is smaller despite the fact that
there are no restrictions from conformal symmetry is that the Burnett 
equations are derived from kinetic theory, which only leads to a subset 
of the terms allowed by the symmetries. 

 We end this section by showing that there are no trace terms in 
$\delta\Pi_{ij}$, despite the fact that second order terms like $g_{ij}
\sigma_{kl}\sigma^{kl}$ are conformal tensors. In Sec.~\ref{sec_kin_0}
we will show that in a non-trivial background the equation of energy
conservation is given be
\beq
\label{e_cons}
\frac{1}{\sqrt{g}} \partial_t \left( \sqrt{g}{\cal E}\right) 
+ \nabla_k \jmath^k_{\epsilon} = -\frac{1}{2}\dot{g}_{ij} \Pi^{ij}\, ,
\eeq
where ${\cal E}={\cal E}_0+\frac{1}{2}\rho V_iV^i$ is the energy
density and $\jmath^i_{\epsilon}=({\cal E}+P)V^k+\delta\jmath^k_\epsilon$
is the energy current. Equ.~(\ref{e_cons}) is conformally invariant if 
\beq 
 {\cal E}_0=\frac{3}{2}P \, , \hspace{1cm}
 g^{ij}\delta\Pi_{ij} = 0 \, ,
\eeq
which shows that the equation of state has to be conformal, and the 
dissipative part of the stress tensor must be traceless. 

%%%%%%%%%%%%%%%%%%%%%%%%%%%%%%%%%%%%%%%%%%%%%%%%%%%%%%%%%%%%%%%%%%%%%%%%%%%
\subsection{Second order terms: Entropy current}
\label{sec_dels_2}
%%%%%%%%%%%%%%%%%%%%%%%%%%%%%%%%%%%%%%%%%%%%%%%%%%%%%%%%%%%%%%%%%%%%%%%%%%%

 In this section we will use the results of Sec.~\ref{sec_inv} to
construct two-derivative conformal vectors that can contribute to 
the entropy current. There is one vectors that can be written as
the comoving derivative of a one-derivative vector 
\beq 
\label{Q_1}
{\cal Q}^1_i = Dq_i\, ,
\eeq
with $[{\cal Q}^{1}_i]=1$. Note that $D\nabla_i\alpha$ is not an 
independent vector, because it can be related to ${\cal Q}^1_i$ and 
the vectors ${\cal Q}^{4,5}_i$ defined below using the hydrodynamic
equation for $D\alpha$. We find four vectors that can be written 
as contractions of one-derivative vectors with one-derivative tensors
\beq 
\label{Q_2345}
{\cal Q}^2_i = q^j\sigma_{ij}\,  , \hspace{0.5cm}
{\cal Q}^3_i = q^j\Omega_{ij}\,  , \hspace{0.5cm}
{\cal Q}^4_i = (\nabla^j\alpha)\sigma_{ij}\,  , \hspace{0.5cm}
{\cal Q}^5_i = (\nabla^j\alpha)\Omega_{ij}\,  , \hspace{0.5cm}
\eeq
with $[{\cal Q}^{2-5}_i]=1$. Finally, there are two vectors that 
can be constructed from the covariant derivative $\nabla_i$ acting
on one-derivative tensors, 
\beq 
\label{Q_67}
{\cal Q}^6_i = \nabla^j\sigma_{ij}\,  , \hspace{0.5cm}
{\cal Q}^7_i = \nabla^j\Omega_{ij}\,  , 
\eeq
with $[{\cal Q}^{6,7}_i]=1$. Again, the number of terms is bigger
than the number of independent parameters in the entropy current 
in the Burnett equation, which is five. 

%%%%%%%%%%%%%%%%%%%%%%%%%%%%%%%%%%%%%%%%%%%%%%%%%%%%%%%%%%%%%%%%%%%%%%%%%%%
\section{Kinetic Theory}
\label{sec_kin} 
%%%%%%%%%%%%%%%%%%%%%%%%%%%%%%%%%%%%%%%%%%%%%%%%%%%%%%%%%%%%%%%%%%%%%%%%%%%

%%%%%%%%%%%%%%%%%%%%%%%%%%%%%%%%%%%%%%%%%%%%%%%%%%%%%%%%%%%%%%%%%%%%%%%%%%%
\subsection{Conservation laws}
\label{sec_kin_0}
%%%%%%%%%%%%%%%%%%%%%%%%%%%%%%%%%%%%%%%%%%%%%%%%%%%%%%%%%%%%%%%%%%%%%%%%%%%

 In this section we will investigate the structure of higher order 
terms starting from the Boltzmann equation. Our motivation is three-fold:
i) We will verify that the results indeed satisfy the constraints derived 
in the previous section; ii) we will show that kinetic theory (within the  
relaxation time approximation) only leads to a subset of the allowed
terms; iii) we will demonstrate that this subset manifestly satisfies 
the second law of thermodynamics. 

 The kinetic equation in a curved background $g_{ij}(x,t)$ can be found 
by starting from the Boltzmann equation in a four-dimensional curved
space \cite{Andreasson:2002vz,Cercignani:2002,Bruhat:2009}
\beq
\label{boltzmann}
\frac{1}{p^{0}}\left(p^{\mu}\frac{\p}{\p x^{\mu}}
  - \Gamma^{i}_{\alpha\beta}p^{\alpha}p^{\beta}\frac{\p}{\p p^{i}}\right)
    f(t,x,p) = C[f]\, ,
\eeq
where $f(t,x,p)$ is the distribution function, $C[f]$ is the collision
term, $\Gamma^\alpha_{\mu\nu}$ is the Christoffel symbol associated 
with the four-dimensional covariant derivative $\nabla_\mu$, $i,j,k$ 
are three-dimensional indices and $\mu,\alpha,\beta$ are four-dimensional 
indices. In the non-relativistic limit \cite{Son:2005rv}
\beq
\label{gam_red}
 \Gamma^{i}_{\al\be}p^{\al}p^{\be}
    =\Gamma^{i}_{jk}p^{j}p^{k} - \frac{1}{m}E^{i}p^{0}p^{0} + 
      \frac{1}{m}g^{il}F_{lk}p^{k}p^{0}+g^{il}\dot{g}_{lk}p^{k}p^{0}\, , 
\eeq
where $p^0\simeq m$, $E^{i}$ is the electric field and $F_{lk}$ is 
the magnetic field introduced in Sect.~\ref{sec_ns}. In the following 
we will drop the Lorentz force (these terms can always be restored 
using the symmetries described in Sect.~\ref{sec_diff}). We get  
\beq
\label{be_nr}
  {\cal D} f(t,x,p) = C[f]\, , 
\eeq
where we have defined the Boltzmann operator
\beq
  {\cal D} = 
  \frac{\p}{\p t} + \frac{p^{k}}{m}\frac{\p}{\p x_{k}}
  - \frac{\Gamma^{i}_{jk}p^{j}p^{k}}{m}\frac{\p}{\p p^{i}}
  - g^{il}\dot{g}_{lk}p^{k}\frac{\p}{\p p^{i}} \, .
\eeq
Consider a collision term that describes elastic two-body collisions
$p_1+p_2\to p_3+p_4$. The symmetries of the collision term imply that 
\beq
 \int\dif\Ga\sqrt{g}\, \chi C[f] = 
    \frac{1}{4}\int\dif\Ga\sqrt{g}\, f 
    C[\chi_{1}+\chi_{2}-\chi_{1'}-\chi_{2'}] \, , 
\eeq
where $\dif\Gamma=d^3p$ and $\chi_i\equiv\chi(p_i)$. This implies that 
moments of the collision term with regard to the collision invariants
$\chi=m,p^i,g_{ij}p^i p^j/(2m)$ must vanish. We will denote averages 
over the momentum distribution as
\beq
\label{def_<a>}
\<A\>=\int\dif\Gamma\,\sqrt{g}f\, A\, .  
\eeq
We define the conserved charges and currents as 
\beq
\label{def_rho_pi_e}
  \rho  = \<m\> \, , \qquad 
  \pi^{k} = \<p^{k}\>\, , \qquad
  {\cal E} = \frac{1}{2m}\left\< g_{ij}p^ip^j \right\> \, , 
\eeq\beq
\label{def_pi_ij}
 \Pi^{ij} = \frac{1}{m} \left\< p^ip^j \right\> \, , \qquad
  \jmath^i_\epsilon = \frac{1}{2m^2}\left\< p^ig_{jk}p^jp^k \right\> \, . 
\eeq
The three collision invariants lead to three conservation laws. The 
first conservation law is obtained by taking a moment of the Boltzmann 
with $\chi=m$. We get 
\bea
0 &=& \int \dif\Gamma\, \sqrt{g}\, m \, {\cal D}f(t,x,p) \nn 
  &=& \frac{\p\l\<m\r\>}{\p t} - \frac{\dot{g}}{2g}\l\<m\r\>
      + \p_{k}\l\<p^{k}\r\> - \frac{\p_{k}g}{2g}\l<p^{k}\r\>
      - \frac{\p}{\p p^{i}}\l\<\Gamma^{i}_{jk}p^{j}p^{k}
      + mg^{il}\dot{g}_{lk}p^{k}\r\>
      + \l\<2\Gamma^{i}_{ji}p^{j} + \frac{m\dot{g}}{g}\r\>  \nn
\label{be_cont}
&=&\frac{\p_{t}\l(\sqrt{g}\rho\r)}{\sqrt{g}}+\del_{k}\pi^{k}\, . 
\eea
The momentum equation follows from taking a moment with $\chi=g_{ij}p^j$,
\beq
\label{be_mom}
0 = \int\dif\Gamma\, \sqrt{g}\, g_{ij}p^j\, {\cal D}f(t,x,p)
  = \frac{\p_{t}\l(\sqrt{g}\pi_{i}\r)}{\sqrt{g}}+\nabla_k\Pi^k_{\;\; i}\, . 
\eeq
and the energy follows from integrating over $\chi=g_{ij}p^ip^j/(2m)$,
\bea
\label{be_e}
0 = \frac{1}{2m}\int\dif\Gamma\, \sqrt{g}\, g_{ij}p^ip^j\, {\cal D}f(t,x,p)
  = \frac{\p_{t}\l(\sqrt{g}{\cal E}\r)}{\sqrt{g}}
         + \nabla_k\jmath^k_\epsilon
         + \frac{1}{2}\dot{g}_{ij}\Pi^{ij} \, .
\eea
This is the result we have used in equ.~(\ref{e_cons}) above. In order 
to determine the form of the currents in terms of the hydrodynamic 
variables we need to know the functional form of the distribution
function near equilibrium. We will see shortly that $f(x,p,t)=f^{(0)}
(g_{ij}c^ic^j)$ with $c_i(x,t)=v_i-V_i(x,t)$ and $v_i=p_i/m$. Then 
\beq
\pi_i = \rho V_i\, , \qquad 
\Pi_{ij} = \rho V_i V_j + Pg_{ij}\, , \qquad
\jmath_{\epsilon}^i = \left({\cal E}_0+P+\frac{1}{2}\rho V_jV^j\right)V^i\, , 
\eeq
with ${\cal E}_0=\frac{1}{2}\langle m(v-V)^2\rangle$ and $P=\frac{2}{3}
{\cal E}_0$. We can now use the thermodynamic relation $d{\cal E}_0=
Tds+\frac{\mu}{m}d\rho$ to determine the entropy equation. We get 
\beq
 \frac{\partial_{t}\l(\sqrt{g}s\r)}{\sqrt{g}}
         + \nabla_k \l(sV^k\r) = 0 \, . 
\eeq
This result can also be derived directly from the Boltzmann equation. 
Computing the moment of the Boltzmann equation with regard to $\log
(f)$ (for classical particles, $\log(f/(1\pm f))$ for Bose/Fermi 
statistics) we obtain equ.~(\ref{h_entr}) with the entropy density 
given by
\beq 
\label{s_be}
s = \int d\Gamma\, \sqrt{g} \, 
 \l[ f\log\l(\frac{1+af}{f}\r)+a\log\l(1+af\r) \r]\, ,
\eeq
and the dissipative function
\beq
\label{R_be}
 \frac{R}{T} =  \int\dif\Gamma\, \sqrt{g}\, 
    \log\left(\frac{f}{1+af}\right) C[f] \, ,
\eeq
with $a=0,\pm 1$ for classical, Bose and Fermi statistics, respectively. 
In the case of classical statistics equ.~(\ref{R_be}) implies that $R$ 
vanishes provided the distribution function is an exponential of the 
collision invariants. The same result holds for quantum statistics, 
with the exponential replaced by Bose-Einstein or Fermi-Dirac 
distribution functions. This implies that
\beq
\label{fzero}
 \fzero(t,x,p) = \fzero\left(
  \frac{g_{ij}(t,x)c^{i}(t,x)c^{j}(t,x)}{2mT(t,x)} - \al(t,x)\right)\, ,
\eeq
where $\fzero$ is the Maxwell, Bose-Einstein, or Fermi-Dirac distribution
and $\alpha=\mu/T$.

%%%%%%%%%%%%%%%%%%%%%%%%%%%%%%%%%%%%%%%%%%%%%%%%%%%%%%%%%%%%%%%%%%%%%%%%%%%
\subsection{First order solution}
\label{sec_kin_1}
%%%%%%%%%%%%%%%%%%%%%%%%%%%%%%%%%%%%%%%%%%%%%%%%%%%%%%%%%%%%%%%%%%%%%%%%%%%

 In this section and the next we will obtain a solution of the Boltzmann
equation to second order in the gradients of the hydrodynamic variables. 
We write the distribution function as 
\beq
f = \fzero+\de f =\fzero+ \fone + \ftwo + \ldots\, , 
\eeq
where $f^{(i)}$ contains terms with $i$ gradients. Since we are 
interested in the structure of the terms that appear in $\delta f$, 
and not in computing the values of transport coefficients for a 
specific theory, we will consider a very simple choice for the 
collision term, the relaxation time approximation
\beq 
\label{C_BGK}
 C[f] = - \frac{\delta f}{\lambda} \, .
\eeq
Conformal invariance implies that $\lambda$ must have conformal
dimension 1. This is physically reasonable, since $\lambda$ has
units of time. The collision term conserves mass, momentum, and 
energy. These conservation laws constrain the form of $\delta f$.
We have
\beq
\label{orth_rel}
\int \dif\chi_c\, \de f = 0\, ,\qquad 
\int \dif\chi_c\, \de f\, c^{i} = 0\, ,\qquad 
\int \dif\chi_c\, \de f\, g_{ij}c^ic^j= 0\, ,
\eeq
where we have defined $\dif\chi_c = \sqrt{g}\,\dif^3c/(2\pi)^3$. The
constraints imply that the correction to the stress tensor due to 
$\de f$ is given by
\beq
\label{del_pi_f}
\de\Pi^{ij} = \int \dif\chi_c \, \de f \, p^ip^j 
  = \int \dif\chi_c\, \de f \, \left(c^ic^j-\frac{1}{3}g^{ij}c^{2}\right)\, 
\eeq
where $c^2\equiv g_{ij}c^ic^j$ and from now on we will use units 
$m\equiv 1$. In order to compute $\fone$ we will follow the procedure 
of Chapman and Enskog \cite{Chapman:1970,Landau:kin} and compute the 
LHS of the Boltzmann equation using the zero'th order solution 
equ.~(\ref{fzero}). We find
\bea
\label{zerolhs_1}
 {\cal D}\fzero &=& -\frac{\fzero(1+a\fzero)}{2T}
    \bigg(c^{i}c^{j}\dot{g}_{ij} + 2\dot{c}^{i}c^{j}g_{ij}
     + p^{k}c^{i}c^{j}\p_{k}g_{ij} + 2p^{k}c^{j}g_{ij}\p_{k}c^{i} \nn
    && \mbox{}
    - 2g_{ik}\Gamma^{i}_{jl}p^{j}p^{l}c^{k} - 2\dot{g}_{kj}p^{j}c^{k}
    - g_{ij}c^{i}c^{j}\left(\p_{t} \log T-p^{k}q_{k}\right)
    - 2T\left(\dot{\al} + p^{k}\p_{k}\al\right)\bigg)\, . 
\eea
Using $\dot c_i=-\p_tV_i$, $\p_j c_i=-\p_jV_i$, and $p_i=c_i+V_i$ 
we obtain
\bea
\label{zerolhs_2}
{\cal D}\fzero &=& \frac{\fzero(1+a\fzero)}{2T}\bigg\{
     \left( 2T\dot{\al}+2TV^{k}\p_{k}\al \right)
    + 2c^{i}\left(\dot{V}_{i}+V^{k}\del_{k}V_{i}+T\p_{i}\al
       -\frac{a_{1}q_{i}}{2}\right) \nn
 && \mbox{}
    + c^{2}\left(\p_{t}\log T-V^{i}q_{i} + 
    \frac{2}{3}\l(\del\cdot V+\frac{\dot{g}}{2g}\r)\right)\nn
 && \mbox{}
    + \l[c^{j}c^{k}-\frac{1}{3}g^{jk}c^2\r]
      \left(\del_{j}V_{k} + \del_{k}V_{j}+\dot{g}_{kj}
    - \frac{2}{3}g_{jk}\l(\del\cdot V+\frac{\dot{g}}{2g}\r)\right) \nn
 && \mbox{}
    - c^{2}c^{k}q_{k}+a_{1}c^{k}q_{k} \bigg\}\, ,
\eea
where, in order to make the constraints equ.~(\ref{orth_rel}) manifest, 
we have added and subtracted the term $a_1c^kq_k$. The first three terms 
in equ.~(\ref{zerolhs_2}) are proportional to the collision invariants 
and must vanish by the equations of motion in order to satisfy the 
constraints. The fourth term, which is proportional to $c_ic_j-\frac{1}{3}
c^2g_{ik}$, automatically satisfies the constraints. The fifth term, 
proportional to $c^2c^k$, satisfies the constraints when combined 
with $a_1c^k$ term. This implies 
\beq
\label{def_a1}
\int\dif\chi_c\, \fzero(1+a\fzero)\, \l( c^4-a_1 c^2 \r)=0\, . 
\eeq
Using the thermodynamic identities derived in App.~\ref{app_thermo}
we find $a_1(t,x)=5P(t,x)/n(t,x)$. For a Maxwell gas the equation of
state is $P=nT$ and $a_1=5T$. The terms proportional to the collision
invariants $1,c_i$ and $c^2$ involve time derivatives of the hydrodynamic
variables $\alpha,T$ and $V_i$. They are easily seen to vanish by the 
Euler equations
\bea
0 &=& \dot{n} + V^{k}\p_{k}n + n\l(\del\cdot V+\frac{\dot{g}}{2g}\r)\, , \\
0 &=& \dot{V}_{i} + V^{k}\del_{k}V_{i} + \frac{\p_{i}P}{n}\, , \\
0 &=& \dot{\cal E}_0 + V^{k}\p_{k}{\cal E}_0 
       + \l( {\cal E}_0+P \r) \l(\del\cdot V + \frac{\dot{g}}{2g}\r) \, .
\eea
We can now solve the first order Boltzmann equation ${\cal D}\fzero=
C[\fzero+\fone]$ for $\fone$. Using equ.~(\ref{C_BGK}) we find
\beq
\label{fone}
\fone = -\frac{\la\fzero(1+a\fzero)}{2T}\l(
    c^{i}c^{j}\si_{ij}- \l( c^2-a_1 \r) c^{k}q_{k}\r)\, , 
\eeq
where $\si_{ij}$ is the shear tensor defined in equ.~(\ref{def_sig}). 
Using the result for $\fone$ we can compute the dissipative corrections
to $\Pi_{ij}$ and $\jmath_s^i$. We find $\delta\Pi_{ij}=-\eta\sigma_{ij}$ 
and $\delta\jmath_s^i = \kappa q^i$. We note that the solution of the 
Boltzmann equation in curved space automatically leads to a conformal 
and diffeomorphism invariant stress tensor and entropy current. The 
shear viscosity and thermal conductivity are given by 
\bea
\label{eta_kin}
\eta   &=& \frac{\lambda}{15T} \int\dif\chi_c\, 
         \fzero \l(1+a\fzero \r) c^4 \, ,\\
\label{kappa_kin}
\kappa &=& \frac{\lambda}{12T^2} \int\dif\chi_c\, 
         \fzero \l(1+a\fzero \r) \l( c^2 - a_1 \r) c^4\, .
\eea 
Using the results in App.~\ref{app_thermo} we can show that $\eta=
\lambda P$ and $\kappa=\frac{\lambda}{12T}(7Q-75P^2/n)$, where $Q=
\langle c^4\rangle$. In the case of a Maxwell gas the result for
the shear viscosity reduces to the familiar form $\eta=\lambda nT$. 
In this limit we also find $\kappa=\frac{5}{2}\lambda nT$, which 
corresponds to a Prandtl ratio ${\it Pr}=c_p\eta/\kappa=1$. Note 
that this result is a consequence of the simple form of the collision 
term in equ.~(\ref{C_BGK}). In a more complete treatment the high 
temperature limit of the Prandtl ratio is ${\it Pr}=\frac{2}{3}$ 
\cite{Braby:2010ec}. Finally, the dissipative function $R$ is given 
by
\bea
R &=& \frac{\eta}{2}\sigma_{ij}\sigma^{ij}
          + \kappa T q_iq^i\, , 
\eea
which is manifestly positive for $\lambda\geq 0$. 

%%%%%%%%%%%%%%%%%%%%%%%%%%%%%%%%%%%%%%%%%%%%%%%%%%%%%%%%%%%%%%%%%%%%%%%%%%%
\subsection{Second order solution}
\label{sec_kin_2}
%%%%%%%%%%%%%%%%%%%%%%%%%%%%%%%%%%%%%%%%%%%%%%%%%%%%%%%%%%%%%%%%%%%%%%%%%%%

 At second order in the derivative expansion the Boltzmann equation
reduces to 
\beq 
\label{be_2}
{\cal D}\fone + \left({\cal D}\fzero+\frac{\fone}{\lambda}\right) 
 = -\frac{\ftwo}{\lambda}\, . 
\eeq
We note that $({\cal D}\fzero+\frac{1}{\lambda}\fone)$ vanishes at 
first order. At second order we have to include gradient terms in 
the equation of motion and this term is not zero. The Boltzmann
operator acting on $\fone$ gives
\bea
{\cal D}\fone &=& -\frac{\la \fzero(1+a\fzero)}{2T}
   \bigg\{\be
    \l( c^ic^j\si_{ij} - \l(c^2-a_1\r) c^kq_k \r)^2  \nn
   && \mbox{}
    + \l( c^{i}c^{j}\si_{ij} - \l(c^{2}-a_1\r) c^{k}q_{k} \r)
      \bigg( D\log(\la) - D\l(\log T\r) \bigg) \nn
   && \mbox{}
    + {\cal D} \l(c^{i}c^{j}\si_{ij} - \l( c^{2}-a_1 \r) c^{k}q_{k} 
     \r)\bigg\}\, , 
\eea
where we have defined $\be=(1+2a\fzero)/(2T)$. We will decompose
this expression as
\beq 
{\cal D}\fone = ({\cal D}f)_{\it orth} 
  + ({\cal D}f)_{\it scal}   + ({\cal D}f)_{\it zm}\, , 
\eeq
where $({\cal D}f)_{\it orth}$ are terms that are orthogonal to the 
zero modes of the collision operator, $({\cal D}f)_{\it scal}$ is
orthogonal to the zero modes but does not contribute to the conserved 
currents $\Pi^{ij}$ and $\jmath^i_\epsilon$, and $({\cal D}f)_{\it zm}$
contains terms that are proportional to the zero modes. This means 
that $({\cal D}f)_{\it zm}$ must cancel against $({\cal D}\fzero+
\frac{1}{\lambda}\fone)$. Before we check this we collect the terms
in $({\cal D}f)_{\it orth}$. We get
{\allowdisplaybreaks
\bea
&({\cal D}f)_{\it orth}  &= -\frac{\la\fzero(1+a\fzero)}{2T}\bigg\{
    \be\l[ c^{i}c^{j}c^{k}c^{l}\si_{ij}\si_{kl}
             -\frac{2}{15} c^{4} \si^{2}\r]
    + a_{1}^{2}\be\l[ c^{i}c^{j}q_{i}q_{j}-\frac{c^{2}q^{2}}{3}\r]
    \nn
    && \mbox{}
    + \be\l[ c^{4}c^{i}c^{j}q_{i}q_{j}-\frac{c^{6}q^{2}}{3}\r]
    - 2a_{1}\be\l[c^{2}c^{i}c^{j}q_{i}q_{j}
    - \frac{c^{4}q^{2}}{3}\r]
    \nn 
    && \mbox{}
    + 2a_{1}\l[\be c^{i}c^{j}c^{k}\si_{ij}q_{k}-c^{i}q_{k}\si_{i}^{k}\r]
    - 2\l[\be c^{2}c^{i}c^{j}c^{k}\si_{ij}q_{k}
                 - \frac{7a_{1}}{5}c^{i}q_{k}\si^{k}_{i}\r]
    \nn
    && \mbox{}
    + \l[c^{i}c^{j}\si_{ij}-c^{2}c^{i}q_{i}+a_{1}c^{i}q_{i}\r]
        \l(D\log(\la)-\p_{t}\log T+V^{k}q_{k}\r)\nn[0.15cm]
    && \mbox{}
    + \l[c^{i}c^{j}c^{k}\si_{ij}q_{k}-\frac{2a_{1}}{5}c^{i}q_{k}\si^{k}_{i}\r]
    - \l[c^{2}c^{i}c^{j}q_{i}q_{j}-\frac{c^{4}q^{2}}{3}\r]
    + \l[a_{1}c^{i}c^{j}q_{i}q_{j}-\frac{a_{1}c^{2}q^{2}}{3}\r]
    \nn
    && \mbox{}
    + \l[c^{i}c^{j}c^{k}-\frac{2a_{1}}{5}c^{i}g^{jk}\r]
              \l(\del_{k}{\si}_{ij}+\si_{ij}q_{k}\r)
    -\l[c^{2}c^{i}c^{j}-\frac{g^{ij}c^{4}}{3}\r]\del_{i}{q}_{j}
    \nn
    && \mbox{}
    + \l[c^{2}c^{k}-a_{1}c^{k}\r]
         \l(q^{i}\del_{k}{V}_{i}+\dot{g}_{ik}q^{i}
          -\dot{q}_{k}-V^{i}\del_{i}{q}_{k}+\frac{2\<\si\>}{3}q_{k}\r)
    \nn
    && \mbox{}
    + \l[c^{i}c^{j}-\frac{g^{ij}c^{2}}{3}\r]
       \bigg(g_{jk}\dot{\si}_{i}^{k}+V^{k}\del_{k}{\si}_{ij}+a_{1}\del_{i}q_{j}
    - \si_{ik}\si^{k}_{j}-\Om_{ik}\si^{k}_{j}-\frac{2\si_{ij}\<\si\>}{3}
    \nn
    \label{d_orth}
    && \hspace{2cm}\mbox{}
    +2 \l(\dot{V}_{i}+V^{k}\del_{k}V_{i}\r)q_{j}+q_{i}\p_{j}a_{1}\bigg)\bigg\}\, .
\eea}
The terms in the square brackets are automatically orthogonal to the 
zero modes. As in the previous section this is achieved by adding and
subtracting terms. Subtraction terms that are proportional to the zero 
modes are collected in $({\cal D}f)_{\it zm}$. The remaining terms are 
collected in $({\cal D}f)_{\it scal}$. We have
{\allowdisplaybreaks
\bea
({\cal D}f)_{\it scal}
    &=& -\frac{\la\fzero(1+a\fzero)}{2T}\bigg\{
     \frac{2\be \si^{2}}{15}\l[ c^{4}-b_{41} c^{2}-b_{40}\r]
     +\frac{\be a_{1}^{2} q^{2}}{3}\l[ c^{2}-b_{21}c^{2}-b_{20}\r]
    \nn
    &&
    +\frac{\be q^{2}}{3}\l[c^{6}-b_{61}c^{2}-b_{60}\r]
    -\frac{2\be a_{1}q^{2}}{3}\l[c^{4}-b_{41}c^{2}-b_{40}\r]
    \nn
    \label{d_scal}
    &&
    -\frac{q^{2}}{3}\l[c^{4}-d_{1}c^{2}-d_{0}\r]
    -\frac{\del_{i}{q}^{i}}{3}\l[c^{4}-d_{1}c^{2}-d_{0}\r]\bigg\}\, . 
\eea}
The constants $b_{20},b_{21},b_{40},b_{41},b_{60},b_{61}$ and $d_1,d_0$ 
are analogous to the constant $a_1$ defined in equ.~(\ref{def_a1}). We 
derive the relevant orthogonality relations in App.~\ref{app_ortho}.
Finally, $({\cal D}f)_{\it zm}$ is given by
{\allowdisplaybreaks
\bea
({\cal D}f)_{\it zm}
    &=& -\frac{\la\fzero(1+a\fzero)}{2T}
       \bigg\{a_{1}Tq^{i}\p_{i}\al+\frac{2b_{40}}{15} \si^{2}
    + \frac{7d_{0}}{6}q^{2}
    - \frac{d_{0}}{3}\del_{i}q^{i}
    \nn
    && \mbox{}
    + c^{2}\l(\frac{\th}{3n}\r)
    \l[- a_{1}Tq^{i}\p_{i}\al -\frac{2b_{40}}{15} \si^{2}
       - \frac{7d_{0}}{6}q^{2}
       + \frac{d_{0}}{3}\del_{i}q^{i} \r]
    \nn
    \label{d_zm}
    && \mbox{}
    + 2c^{i}\l[\frac{\si_{i}^{j}\p_{j}P}{n}
    + \frac{P}{n}\del_{k}\si^{k}_{i}\r]
    \bigg\}\, , 
\eea}
where we have used the Euler equations to eliminate time derivatives 
of the hydrodynamic variables. This is consistent at this order 
in the derivative expansion. The parameter $\theta$ is defined in 
App.~\ref{app_thermo} (for a Maxwell gas $\theta=n/T$). We now use 
the Navier-Stokes equations
\bea
0&=& 2T\dot{\al} + 2TV^{k}\p_{k}\al  \nn
  && \mbox{}
     + \la\l[a_{1}Tq^{i}\p_{i}\al+\frac{2b_{40}}{15} \si^{2}
     + \frac{7d_{0}}{6}q^{2}
     - \frac{d_{0}}{3}\del_{i}q^{i}\r] \, , \\
0&=& \dot{V}_{i} + V^{k}\del_{k}V_{i} + T\p_{i}\al
     - \frac{a_{1}q_{i}}{2}+\la\l[\frac{\si_{i}^{j}\p_{j}P}{n}
     + \frac{P}{n}\del_{k}\si^{k}_{i}\r] \, ,  \\
0&=& \p_{t}\log T-V^{i}q_{i} 
        + \frac{2}{3}\l(\del\cdot V + \frac{\dot{g}}{2g}\r)\nn
 && \mbox{}
    + \la\l(\frac{\th}{3n}\r)\l[- a_{1}Tq^{i}\p_{i}\al
      - \frac{2b_{40}}{15} \si^{2}
      - \frac{7d_{0}}{6}q^{2}
      + \frac{d_{0}}{3}\del_{i}q^{i}
    \r] \, , 
\eea
to confirm that $({\cal D}f)_{\it zm}+({\cal D}\fzero+\frac{1}{\lambda}
\fone)=0$. Solving equ.~(\ref{be_2}) for $f^{(2)}$ we get
\beq
\label{ftwo}
\ftwo =-\la \l[ ({\cal D}f)_{\it orth}+({\cal D}f)_{\it scal} \r]\, . 
\eeq
where $({\cal D}f)_{\it orth}$ and $({\cal D}f)_{\it scal}$ are defined
in equ.~(\ref{d_orth}) and (\ref{d_scal}). We note that $({\cal D}
f)_{\it orth}$ depends on $D\log(\lambda)$. This term is necessary in
order to respect conformal invariance. Dimensional analysis implies 
that $\lambda=T^{-1}h(\alpha)$ for some undetermined function $h$. 
Using the Euler equation we can then show that $D\log(\lambda)=
\frac{2}{3}\langle\sigma\rangle$.

%%%%%%%%%%%%%%%%%%%%%%%%%%%%%%%%%%%%%%%%%%%%%%%%%%%%%%%%%%%%%%%%%%%%%%%%%%%
\subsection{Stress tensor and entropy current at second order}
\label{sec_del_pi_2_kin}
%%%%%%%%%%%%%%%%%%%%%%%%%%%%%%%%%%%%%%%%%%%%%%%%%%%%%%%%%%%%%%%%%%%%%%%%%%%

 Using the result for $\ftwo$ obtained in the previous section we 
can compute the dissipative correction to the stress tensor at 
second order in the gradient expansion. As discussed above only
$({\cal D}f)_{\it orth}$ contributes to the conserved currents. 
We will write $\ftwo_{\it orth}=-\la({\cal D}f)_{\it orth}$. We
can simplify $\ftwo_{\it orth}$ by combining some terms, and by 
using the Euler equation. We find
{\allowdisplaybreaks
\bea
\ftwo_{orth}
 &=& \frac{\la^2\fzero(1+a\fzero)}{2T}\bigg\{
    \be\l[ c^{i}c^{j}c^{k}c^{l}\si_{ij}\si_{kl}-\frac{2}{15}{c}^{4}{\si}^{2}\r]
    \nn
    &&\mbox{}
    +\l({c}^{2}-a_{1}\r)\l(\be{c}^{2}-a_{1}\be-1\r)
         \l[ c^{i}c^{j}q_{i}q_{j}-\frac{{c}^{2}{q}^{2}}{3}\r]
    \nn
    &&\mbox{}
    + \l[c^{i}c^{j}\si_{ij}-c^{2}c^{i}q_{i}+a_{1}c^{i}q_{i}\r]
           \frac{2}{3}\<\sigma\>
    + 2\l(a_{1}\be+1-\be{c}^{2}\r)c^{i}c^{j}c^{k}\si_{ij}q_{k}
    \nn[0.2cm]
    &&\mbox{}
    + \l[c^{i}c^{j}c^{k}-\frac{2a_{1}}{5}c^{i}g^{jk}\r]\del_{k}{\si}_{ij}
    + \l[{c}^{2}c^{k}-a_{1}c^{k}\r]
        \l(q^{i}\del_{k}{V}_{i}+\dot{g}_{ik}q^{i}
            -\dot{q}_{k}-V^{i}\del_{i}{q}_{k}\r) 
    \nn
    &&\mbox{}
    +\l[c^{i}c^{j}-\frac{g^{ij}{c}^{2}}{3}\r]
     \Big(g_{jk}\dot{\si}_{i}^{k}+V^{k}\del_{k}{\si}_{ij}
        - \si_{ik}\si^{k}_{j} - \Om_{ik}\si^{k}_{j} 
    \nn
    \label{d_orth_2}
    && \hspace{3.25cm}\mbox{}
        + \frac{(3n^{2}-5P\th) T}{n^{2}}q_{i}\p_{j}\al
        - \l({c}^{2}-a_{1}\r)\del_{i}q_{j}\Big) \bigg\}\, . 
\eea}
The stress tensor can be determined using equ.~(\ref{del_pi_f}). The
result is 
\bea
\de\Pi_{(2)}^{ij}
&=& \la\eta\l(\si^{ik}\si_{k}^{j} - \frac{g^{ij}}{3}\si^{2}\r)
 - \frac{4\la^2}{5}\del^{(i}\l(\frac{\ka T}{\la} q^{j)}\r)
 + \frac{4\la^2 g^{ij}}{15}\del_{k}\l(\frac{\ka T}{\la} q^{k}\r)  \nn
& &\mbox{}
 + \la\eta\l(g^{jk}\dot{\si}^{i}_{k}+V^{k}\del_{k}{\si}^{ij}
 + \frac{2}{3}\langle\sigma\rangle \si^{ij}-\Om^{(ik}\si_{k}^{j)}\r)\, .
\eea
This result is in agreement with the general form of $\delta\Pi^{ij}$
derived in Sect.~\ref{sec_delpi_2}. We note that some of the terms 
allowed by symmetry vanish in the kinetic theory calculation. This 
includes the $\Omega^{i}_{\;\; k}\Omega_{}^{jk}$ structure and the 
$\nabla_iP\nabla_jP$ and $\nabla_i\nabla_j P$ terms. The kinetic 
theory also does not give a pure curvature term proportional to 
$R_{ij}$. We also note that the coefficient of the $g^{jk}\dot{\si}^{i}_{k}$
term can be written as $\tau_R\eta$, where $\tau_R$ is the relaxation 
time for the dissipative stresses. Our result shows that $\tau_R
=\lambda=\eta/P$. This result is in agreement with the calculation 
of spectral function of the stress tensor in \cite{Chao:2010tk}, 
and with previous work on relaxation effects in dilute Bose and Fermi
gases \cite{Nikuni:2004,Bruun:2007}.

 Finally, we compute dissipative corrections to the entropy current
and the dissipative function. We will use the expression for $\jmath_s^i$
and $R$ that are obtained by taking moments of the Boltzmann equation
with $\log(f/(1+af))$, see Sect.~\ref{sec_kin_0}. As a consequence
of the H-theorem these equations automatically satisfy the second law
of thermodynamics. This is no longer the case if the result is expanded 
in derivatives of the thermodynamic variables. The distribution function 
at second order in the gradient expansion determines $R$ to third order 
in gradients, and this expression is not manifestly positive. It is 
possible, however, to add certain fourth order terms and obtain a 
manifestly positive result. This is the strategy we have adopted. We 
find
\bea
\jmath_s^i &=&  \ka q^{i}
 - \la\, \bigg(\frac{27\ka q_{k}\si^{ki} }{10}
   -\frac{2\ka }{5}\del_{k}\si^{ki}
   -\frac{\ka q_{k}\Om^{ik}}{2}+\frac{\ka q^{i}\<\si\>}{3} \nn
 \label{j_s_2}
 &&\hspace{4cm}\mbox{}
  + \ka \dot{q}^{i}+\frac{\ka g^{ik}\dot{g}_{kj}q^{j}}{2}
  + \ka V^{k}\del_{k}{q}^{i}
   \bigg)\, ,
\eea
and
\bea
 \frac{R}{T}
 &=&\frac{\eta}{2}\l(\si - \la\l[ \frac{1}{2}\si\si - \si\Om
     + \frac{2}{3}\<\si\>\si + \dot{\si} + V^{k}\del_{k}\si\r]\r)^{2}\nn
   \label{R_2}
 && \mbox{}
     + \ka T\l(q - \la\l[ \frac{27 q\si}{20} - \frac{q\Om}{2}
     + \frac{q\<\si\>}{3} +  \dot{q} +  V^{k}\del_{k}q\r]\r)^{2}\, .
\eea
In equ.~(\ref{R_2}) we have suppressed tensor indices. Contractions
are defined as $\si\si = \si^{}_{ik}\si^{k}_{\;j}$, $\si\Om=\si_{ik}
\Om^{k}_{\; j}$ etc., and $(\si)^2=\si_{ij}\si^{ij}$. We note that 
the dissipative function is manifestly positive provided the first 
order relations $\eta\geq 0$ and $\kappa\geq 0$ are satisfied, and 
that the entropy current is of the form discussed in Sect.~\ref{sec_dels_2}.  

%%%%%%%%%%%%%%%%%%%%%%%%%%%%%%%%%%%%%%%%%%%%%%%%%%%%%%%%%%%%%%%%%%%%%%%%%%%
\section{Outlook and conclusions}
\label{sec_out}
%%%%%%%%%%%%%%%%%%%%%%%%%%%%%%%%%%%%%%%%%%%%%%%%%%%%%%%%%%%%%%%%%%%%%%%%%%%

 In this work we have studied the constraints imposed by conformal
invariance on the form of the hydrodynamic equations at second
order in the gradient expansion. We find that the most general
form of the stress tensor is given by 
\bea 
\delta\Pi_{ij} &=& -\eta\sigma_{ij}
   + \eta\tau_R\l(
    g_{ik}\dot\sigma^{k}_{\; j} + V^k\nabla_k \sigma_{ij}
    + \frac{2}{3} \langle \sigma\rangle \sigma_{ij} \r) 
    + \la_1 \sigma_{\langle i}^{\;\;\; k}\sigma^{}_{j\rangle k} 
    + \la_2 \sigma_{\langle i}^{\;\;\; k}\Omega^{}_{j\rangle k}\nn
   && \mbox{} 
    + \la_3 \Omega_{\langle i}^{\;\;\; k}\Omega^{}_{j\rangle k}  
    + \ga_1 \nabla_{\langle i}T\nabla_{j\rangle}T
    + \ga_2 \nabla_{\langle i}P\nabla_{j\rangle}P
    + \ga_3 \nabla_{\langle i}T\nabla_{j\rangle}P  \nn
   \label{del_pi_fin}
   && \mbox{}
    + \ga_4 \nabla_{\langle i}\nabla_{j\rangle}T 
    + \ga_5 \nabla_{\langle i}\nabla_{j\rangle}P
    + \ka_R  R_{\langle ij\rangle}\, , 
\eea
where we have suppressed terms involving the gauge fields. 
Conformal symmetry constrains the form of the comoving derivative
of $\sigma_{ij}$ and eliminates possible trace terms like $g_{ij}
\sigma_{kl}\sigma^{kl}$. For phenomenological applications it is 
useful to rewrite the second order equations as a relaxation equation
for the viscous stress $\pi_{ij}\equiv\delta\Pi_{ij}$. For this 
purpose we use the first order relation $\pi_{ij}=-\eta\sigma_{ij}$ 
and rewrite equ.~(\ref{del_pi_fin}) as 
\beq 
\label{pi_rel}
\pi_{ij} = -\eta\sigma_{ij}
   - \tau_R\l(
    g_{ik}\dot\pi^{k}_{\; j} + V^k\nabla_k \pi_{ij}
    + \frac{5}{3} \langle \sigma\rangle \pi_{ij} \r) 
    + \la_1 \pi_{\langle i}^{\;\;\; k}\pi^{}_{j\rangle k} 
    + \ldots \, ,
\eeq
where $\ldots$ refers to the terms proportional to $\lambda_{2,3},
\gamma_i$ and $\kappa_R$. In deriving equ.~(\ref{pi_rel}) we have 
also used $\eta(n,T)=n f(\alpha)$, where $f$ is a function of 
$\alpha$, and the Euler equations for $n$ and $\alpha$. 
Equ.~(\ref{pi_rel}) has the same structure as the second 
order equations considered by Israel and Stewart \cite{Israel:1979wp},
but the coefficient $\frac{5}{3}$ is specific to the non-relativistic
conformal case. 

 We have checked the conformal constraints by computing the dissipative 
contribution to the stress tensor, the entropy current, and the 
dissipative function in kinetic theory. We find that $\delta 
\Pi_{ij}$ is indeed of the form given in equ.~(\ref{del_pi_fin}), 
but that some terms allowed by the symmetries do not appear in 
kinetic theory. The relaxation time for the dissipative stresses 
is given by $\tau_R=\eta/P$. The entropy current and the dissipative 
function also satisfy the conformal constraints. As in the case of 
the stress tensor, not all possible terms appear. The relaxation 
time for the entropy current is equal to the viscous relaxation time. 

 There are a number of issues that we have not addressed in this
paper. We have studied the constraints that arise from conformal
symmetry and Galilean invariance, but we have not investigated 
the conditions that arise from the second law of thermodynamics. 
These conditions should have the form of inequalities for the 
transport coefficients, analogous to the first order relations 
$\eta\geq 0$ and $\kappa\geq 0$. In the kinetic theory calculation
the second law is satisfied automatically, but in the simple
model considered here the transport coefficients are all governed 
by a single relaxation time $\lambda$. We have also not attempted
to derive Kubo relations for the coefficients $\tau_R$, $\lambda_i$,
and $\gamma_i$. The analogous Kubo formulas in the relativistic
case were obtained in \cite{Moore:2010bu}. Finally, it would be 
interesting to compute the spectral functions for the stress
tensor and the energy current at second order in the derivative 
expansion. These results would be useful in connection with 
attempts to compute transport coefficients using Quantum Monte
Carlo methods \cite{Meyer:2011gj}.

Acknowledgments: This work was supported in parts by the US
Department of Energy grant DE-FG02-03ER41260. We thank Dam 
Son for a useful discussion. 

\appendix

%%%%%%%%%%%%%%%%%%%%%%%%%%%%%%%%%%%%%%%%%%%%%%%%%%%%%%%%%%%%%%%%%%%%%%%%%%%
\section{Thermodynamic relations}
\label{app_thermo}
%%%%%%%%%%%%%%%%%%%%%%%%%%%%%%%%%%%%%%%%%%%%%%%%%%%%%%%%%%%%%%%%%%%%%%%%%%%

 In this appendix we collect a number of thermodynamic relations.
In kinetic theory the number density and the pressure are determined
by equ.~(\ref{def_rho_pi_e}) and (\ref{def_pi_ij}). We have
\bea
n(T,\al)&=&\;\;\int\dif\chi_c\,
          \fzero \left(\frac{c^{2}}{2T}-\al\right) \, ,\\
P(T,\al)&=&\frac{1}{3}\int\dif\chi_c\,c^2
          \fzero \left(\frac{c^{2}}{2T}-\al\right)\, ,
\eea
where $\dif\chi_c$ is defined below equ.~(\ref{orth_rel}) and we 
have set $m=1$. We also define
\beq
Q(T,\al) = \;\;\int\dif\chi_c\,c^4
          \fzero \left(\frac{c^{2}}{2T}-\al\right)\, .
\eeq
We can now compute various differentials. Differentials
of $n$ are given by 
\bea
\frac{\p n(T,\al)}{\p T}\bigg|_{\al}
 &=& \p_{T}\l\{\int\dif\chi_c
   \fzero \left(\frac{c^{2}}{2T}-\al\right)\r\}
  = \frac{1}{2T^{2}}\int\dif\chi_c\, c^{2}\fzero(1+a\fzero)
  = \frac{3n}{2T}\, , \\
\frac{\p n(T,\al)}{\p\al}\bigg|_{T}
 &=& \p_{\al}\l\{\int\dif\chi_c\,
   \fzero\left(\frac{c^{2}}{2T}-\al\right)\r\}
  = \int\dif\chi_c\, \fzero(1+a\fzero)
  = \th T\, , 
\eea
where $a=0,\pm 1$ corresponds to Maxwell, Bose-Einstein and Fermi-Dirac
distribution functions, respectively. From the above relations we find
\beq
\dif n= \th T\dif\al + \frac{3n}{2}\dif\l(\log T\r)\, . 
\eeq
Differentials of the pressure can be computed in the same way. We get
\bea
\frac{\p P(T,\al)}{\p T}\bigg|_{\al}
 &=& \frac{1}{3}\p_{T}\l\{\int\dif\chi_c\, c^{2}
   \fzero\left(\frac{c^{2}}{2T}-\al\right)\r\}
  = \frac{5P}{2T}\, , \\
\frac{\p P(T,\al)}{\p \al}\bigg|_{T}
 &=&\frac{1}{3}\p_{\al}\l\{\int\dif\chi_c\,  c^{2}
   \fzero\left(\frac{c^{2}}{2T}-\al\right)\r\}
 = nT\, ,
\eea
and
\beq
\dif P = nT\dif\al + \frac{5P}{2}\dif\l(\log T\r) \, ,\qquad
\dif {\cal E}_0 =\frac{3nT}{2}\dif\al + \frac{15P}{4}\dif\l(\log T\r)\, . 
\eeq
From ${\cal E}_0+P=\mu n+Ts$ we obtain
\bea
\frac{\p s(T,\al)}{\p T}\bigg|_{\al}
 &=& - \frac{5P}{2nT^{2}} 
     + \frac{5}{2nT}\frac{\p P(T,\al)}{\p T}\bigg|_{\al}
     - \frac{5P}{2n^{2}T}\frac{\p n(T,\al)}{\p T}\bigg|_{\al}
  = 0\, ,  \\
\frac{\p s(T,\al)}{\p \al}\bigg|_{T}
 &=& - 1 
     + \frac{5}{2nT}\frac{\p P(T,\al)}{\p \al}\bigg|_{T}
     - \frac{5P}{2n^{2}T}\frac{\p n(T,\al)}{\p \al}\bigg|_{T}
  = \frac{3n^{2}-5P\th}{2n^{2}}\, ,
\eea
and
\beq
\dif s = \frac{3n^{2}-5P\th}{2n^{2}}\dif\al\, .
\eeq
For a Maxwell gas this relation becomes $\dif s=-\dif\al$. The
differential of $Q$ is given by
\beq 
\dif Q = 15PT\dif\al + \frac{7Q}{2}\dif\l(\log T\r)\, . 
\eeq

%%%%%%%%%%%%%%%%%%%%%%%%%%%%%%%%%%%%%%%%%%%%%%%%%%%%%%%%%%%%%%%%%%%%%%%%%%%
\section{Orthogonality relations}
\label{app_ortho}
%%%%%%%%%%%%%%%%%%%%%%%%%%%%%%%%%%%%%%%%%%%%%%%%%%%%%%%%%%%%%%%%%%%%%%%%%%%

 In this appendix we collect the orthogonality relations that determine
the coefficients $b_{20},b_{21},b_{40},b_{41},b_{60},b_{61}$ and $d_1,d_0$
introduced in Sec.~\ref{sec_kin_2}. The coefficients $b_{20},b_{21}$ 
are defined by
\bea
0 &=& \int\dif\chi_c\, \fzero(1+a\fzero)\l(\frac{1+2a\fzero}{2T}
     c^{2}-b_{21}c^{2}-b_{20}\r) \, , \\
0 &=& \int\dif\chi_c\, \fzero(1+a\fzero)\l(\frac{1+2a\fzero}{2T}
     c^{4}-b_{21}c^{4}-b_{20}c^{2}\r)\, . 
\eea
These relations imply
\beq
     b_{21} = \frac{-n\th}{3n^{2}-5P\th},\qquad 
     b_{20} = \frac{15(n^{2}-P\th)}{2(3n^{2}-5P\th)} \, . 
\eeq
The remaining coefficients are determined analogously. We have
\bea
0 &=& \int\dif\chi_c\, \fzero(1+a\fzero)\l(\frac{1+2a\fzero}{2T}
     c^{4}-b_{41}c^{2}-b_{40}\r)  \, , \\
0 &=& \int\dif\chi_c\, \fzero(1+a\fzero)\l(\frac{1+2a\fzero}{2T}
     c^{6}-b_{41}c^{4}-b_{40}c^{2}\r) \, ,\\
  &\Rightarrow&
      b_{41} = \frac{5(3n^{2}-7P\th)}{2(3n^{2}-5P\th)}\, ,\qquad 
      b_{40} = \frac{15Pn}{3n^{2}-5P\th}\, , \\[0.1cm]
0 &=& \int\dif\chi_c\, \fzero(1+a\fzero)\l(\frac{1+2a\fzero}{2T}
     c^{6}-b_{61}c^{2}-b_{60}\r) \, ,\\
0 &=& \int\dif\chi_c\, \fzero(1+a\fzero)\l(\frac{1+2a\fzero}{2T}
     c^{8}-b_{61}c^{4}-b_{60}c^{2}\r) \, ,  \\
  &\Rightarrow&
      b_{61}=\frac{21(5nP-Q\th)}{2(3n^{2}-5P\th)}\, ,\qquad 
      b_{60}=\frac{21(3nQ-25P^{2})}{2(3n^{2}-5P\th)}\, , \\[0.1cm]
0 &=& \int\dif\chi_c\, \fzero(1+a\fzero)\l(
     c^{4}-d_{1}c^{2}-d_{0}\r)\, ,  \\
0 &=& \int\dif\chi_c\, \fzero(1+a\fzero)\l(
     c^{6}-d_{1}c^{4}-d_{0}c^{2}\r) \, , \\
  &\Rightarrow& 
       d_{1} = \frac{45nP-7Q\th}{3(3n^{2}-5P\th)}\, ,\qquad 
       d_{0} = \frac{7nQ-75P^{2}}{3n^{2}-5P\th}\, . 
\eea
where the thermodynamic variables $\th,n,P,Q$ are defined in 
the previous section.

\end{document}